\documentstyle[prl,aps,preprint,fps]{revtex}

\newcommand{\be}{\begin{eqnarray}} \newcommand{\ee}{\end{eqnarray}}
\newcommand{\lb}[1]{\label{#1}} \newcommand{\e}{{\bf\varepsilon}}
 \newcommand{\La}{{\Lambda}}
\newcommand{\C}{\overline P + \bar\kappa}
\newcommand{\rf}[1]{~(\ref{#1})} 
\newcommand{\abs}[1]{\left| #1\right|}
\newcommand{\ct}[1]{$^{\cite{#1}}$}
   
\begin{document}
\setcounter{page}{0}
\def\footnoterule{\kern-3pt \hrule width\hsize \kern3pt}
\tighten
% TITLE
\title{Quark States near a Threshold\\ and\\
       the Unstable H-dibaryon\thanks
{This work is supported in part by funds provided by the U.S.
Department of Energy (D.O.E.) under cooperative 
research agreement \#DF-FC02-94ER40818.}}

\author{S.~V.~Bashinsky and R.~L.~Jaffe}

\address{{~}\\Center for Theoretical Physics \\
Laboratory for Nuclear Science \\
and Department of Physics \\
Massachusetts Institute of Technology \\
Cambridge, Massachusetts 02139 \\
{~}}

\date{MIT-CTP-2643,~hep-ph/9705407. {~~~~~} May 1997}
\maketitle

\thispagestyle{empty}

\begin{abstract}

We consider the interplay of a quark state and a hadronic threshold  
in the framework of the $P$-matrix formalism, which is reviewed and extended 
for use together with conventional methods of computing quark-gluon dynamics.
We provide a quantitative dynamical interpretation of the reduced $R$ or $K$ 
matrices and their poles that suggests a natural classification of threshold 
phenomena. At a threshold with a quark state close to it up to three 
$S$-matrix poles can be found.
The scattering amplitudes for the corresponding cases are discussed.  
 Our analysis is applied to make an outlook for 
experimental observation of the doubly strange $H$-dibaryon if
it is not stable to strong decays. 
  
\end{abstract}

\newpage

\section{Introduction and Summary}
 
A resonance shape can be dramatically distorted if one of its decay 
channels has a threshold within the resonance width.  A tiny variation 
of coupling strength may lead to a wide spectrum of physical phenomena 
such as a slightly bound or a virtual state, a ``shoulder'', or a resonance.  
All these effects are of kinematic origin.  We will show that the underlying 
quark-gluon dynamics can be isolated and quantitatively estimated in a smooth 
way which is unaffected by such kinematic cataclysms.

There is little doubt that far from threshold singularities, narrow and 
dramatic effects in scattering amplitudes are to be identified with 
quasi-stable states of QCD.  Little sophistication is required to connect 
the $\rho(770)$ with $\bar u u {-} \bar d d$ or the $\phi(1020)$ with 
$\bar s s$.  
However, great care must be used when attempting to assign a fundamental QCD 
interpretation to broad effects like most of those seen in meson-meson 
scattering above 1 GeV or to striking effects like the $f_0(980)$ and 
$a_0(980)$ that lie near thresholds (in this case $K\bar K$). 
Identification of many objects of great interest --- exotics, hybrids, 
glueballs,  quasi-molecular 
states, {\it etc.\/} --- requires us to relate low energy 
scattering consistently to microscopic quark-gluon dynamics.

We study hadron-hadron scattering at small kinetic energy, where 
non-relativistic methods suffice.  This is an old problem, but there is no 
general agreement on how to associate quark-gluon ``states'' with effects 
seen in low energy scattering.
One of the most popular phenomenological tools is the 
$K$-matrix parameterization$^{\cite{Levisetti}}$ and its pole analysis. 
The $K$-matrix emerges naturally when the interaction is localized
at distances small with respect to the de Broglie wavelength 
of the scattered particles. 
For a single channel and an energy close enough to the threshold, 
the conditions for a
$K$-matrix analysis might seem to be met.  However, in the real world
hadron-hadron systems with small relative momentum are often strongly
coupled to other open or closed channels where the wavelength does not
exceed the range of interaction. In this case results obtained from 
solving microscopic quark dynamics must not be directly associated 
with the many-channel $K$-matrix. In place of the $K$-matrix, we will 
argue that the $P$-matrix formalism$^{\cite{P-orig}}$  is more suitable for 
this purpose. We will also show that the {\it reduced} $K$-matrix does
allow certain dynamical interpretation.

The theoretical part of our paper, Sections~2 and~3, 
consists of two general divisions. Sec.~2 is concerned
with micro-dynamics on the hadron-size scale and the $P$-matrix formalism.
The following Sec.~3 deals with observable objects such as 
$S$-matrix and cross-sections. Some of the results and the organization of 
these sections are outlined in the rest of the introduction. 
In Section~4 we discuss the phenomenological implications of 
our work for the search for a doubly strange 
$H$-dibaryon if this six-quark state is not bound.

In Section~2a we review and extend the $P$-matrix formalism.
$P$ is defined, similar to $K$, as an algebraic transform of the 
$S$-matrix but it involves an additional parameter $b$:
\be
P(\e,b)=i{~}\sqrt{k}{~}\frac{e^{ikb}S(\e)e^{ikb}+1}
{e^{ikb}S(\e)e^{ikb}-1}{~}\sqrt{k}{~},
\lb{PS}
\ee
where we consider a multichannel $s$-wave with the total energy $\e$.
If for relative distance $r{>}b$ the hadrons do not interact or 
the interaction is simple enough to be 
described with a potential, the $P$-matrix generalizes the logarithmic
derivative  of the wave function at $r{=}b$ (see Section~2a for details). 
Then $P$ is fully determined by the dynamics in the inner domain 
$r{<}b$.  The poles of $P(\e,b)$, 
or {\it primitives}, play an important role. 
Their positions and residues are related to the spectrum of a
system confined in a hypothetical spherical shell with a radius 
depending on $b$. Such boundary conditions are used in the bag model and 
could be simulated on a lattice.  In Section~2b we illustrate the 
$P$-matrix calculation taking the bag model as an example. 
The issue of flavor symmetry is addressed.
We show that $P_{ij}(\e,b)$ reflects this symmetry provided the shell is 
sufficiently small. Then the flavor projection of the quark-bag states 
onto a two-hadron state determines the corresponding projection 
of the $P$-pole residues.

In Section~3a we reconstruct the $S$-matrix from the $P$-matrix 
and see that a pole in $P(\e)$ gives a resonance-like term in $S$.
Given an arbitrary background scattering, $\overline S(\e)$, the 
$S$-matrix will be written as
\be
S_{ij}=\overline S_{ij}-i\chi_{i}{~}\frac{1}{\e-\e_r+
  i{~}{\displaystyle\frac{\gamma}{2}}}
       {~}\chi_{j}{~}.
\lb{Sintro}
\ee
where $\chi_{i}(\e)$, $\e_r(\e)$, and $\gamma(\e)$ are specified by
$\overline S(\e)$ and the $P$-matrix poles. 
The unitarity of the $S$-matrix is preserved automatically. 
For a narrow $P$ pole eq.\rf{Sintro} is, in fact,
a Breit-Wigner resonance but in general the energy dependence in 
$\chi_{i}$, $\e_r$, and $\gamma$ allows a broad spectrum of physical 
phenomena. We discuss those that arise when a pole 
in $P(\e)$ (a primitive) occurs near a hadronic threshold in the following
Section~3b. We start from a review of the analytical structure of the
many-channel $S$-matrix
at a threshold. It will be shown that the pole in $P$-matrix close to the 
threshold 
gives rise to up to {\it two} poles in $S(\e)$ that influence the scattering 
and another, in general {\it third}, pole may appear in the $S$-matrix 
due to potential, or background, scattering. After these preliminaries 
we consider the inverse logarithmic derivative of the 
two-hadron wave function at $r{=}b$ in the channel with the threshold. 
This quantity is named as the reduced $R$-matrix, $R^{(red)}$,
after Wigner and Eisenbud and is approximately equal to the reduced $K$-matrix
close to the threshold. We obtain a formula for $R^{(red)}$ that 
will appear to have a transparent dynamical interpretation in which
the quark state corresponds to the pole in $R^{(red)}(\e)$ while 
the potential scattering specifies its regular part.  
This result to some degree justifies the interpretation of the poles 
in the $K$-matrix as manifestations of 
narrow quark states.\footnote{
Let us emphasize that this interpretation is possible only for the 
{\it reduced} $K$ or $R$ matrices and looks controversial 
at first glance since a pole in $K^{(red)}$ or $R^{(red)}$ occurs when the 
derivative of the hadron wave function vanishes, 
$\psi'(r){=}0$, while the boundary
conditions for the quark-bag states are just the opposite: $\psi(r){=}0$.
}
With the formalism developed we classify the possible threshold effects and 
consider scattering amplitudes and cross-sections for the corresponding cases. 
Many of them are illustrated by existing physical systems.

 In the Section~4 the previous results are used to explore 
the experimental manifestations of an unbound six-quark 
$H$-dibaryon. In the Appendix we estimate parameters giving
the width of this state and the hadronic shift in the $H$
mass due to the influence of open channels.
\section{The P-matrix}

This method of analyzing two-body reactions was proposed by Jaffe and Low 
in 1979 in order to test the spectroscopic predictions of quark models
especially as they relate to exotic ({\it e.g.\/} multi-quark) states. 
The formalism was initially developed in the context of the bag model, where 
quarks are confined by a scalar vacuum pressure. However, it applies to any 
model in which quark and gluon eigenstates are studied without considering 
their coupling to decay channels.
First, we briefly review this formalism and 
also present new arguments that  give a further insight into the
connection between low-energy scattering and  quark model speculations. 
In the next subsection we quantitatively estimate  the parameters of the 
$P$-matrix from the quark-bag model for various two-hadron systems. 

\subsection{Formalism}

At low kinetic energies hadron-hadron scattering may be described
by non-relativistic kinematics. Restricting our attention to zero total
spin and zero angular momentum, $S{=}L{=}0$,
we factor out the center-of-mass motion and consider the wave function of 
an $n$-channel two-hadron system in the relative coordinate $r$.
For a given value of a spatial parameter~$b$, a definite energy~$\e$, 
and $r$ greater than the interaction radius, the most general form of the
wave function is  
\be
\psi_i(r_i)=\sum_{j=1}^n\left\{\cos[k_i(r_i-b)]{~}\delta_{ij} 
+\frac{\sin[k_i(r_i-b)]}{k_i}{~}P_{ij} \right\} A_j, 
\lb{defP}
\ee
where $i=1,{~}...{~},n$ labels the channel and the  $\{A_j\}$ 
are some amplitudes.

The matrix $P_{ij}$ generalizes the logarithmic derivative of $\psi(r)$ 
for the case of many channels. Comparing eq.~(\ref{defP}) to the usual 
$S$-matrix parameterization of the scattering wave function and 
assuming that the reduced two-hadron masses, $m$, are almost the same in
all $n$ channels, we find that 
$P$ and $S$-matrices are simply related as$^{\cite{P-orig}}$:
\be 
S=e^{-ikb} \frac{{ \frac{1}{\sqrt{k}}} P
 {\frac{1}{\sqrt{k}}} +i}{{\frac{1}{\sqrt{k}}}P
 {\frac{1}{\sqrt{k}}}-i} e^{-ikb}{~}.
\lb{SP}
\ee
After this matrix equation is resolved with respect 
to $P$, one arrives at eq.~(\ref{PS}).
The unitarity of $S$ requires the $P$-matrix to be hermitian\footnote{
At the energy when only the first $m<n$ channels are open, only the 
upper-left $m\times m$ sub-matrix of $S_{ij}$  is unitary.
Nevertheless, the whole $n\times n$ $P$-matrix is hermitian for all energies.}.
If the interaction is time
reversal invariant, then $P$ is also real. 
$P$ depends on $b$ according to the equation$^{\cite{P-orig}}$

\be
\frac{\partial P}{\partial b}=-P^2-k^2{~}.
\lb{dP/db}
\ee
  
For the present we treat $b$ as a free parameter. Suppose for a moment 
that the value of~$b$ is large enough so that there is no interaction 
between hadrons for $r{\geq} b$. If for the energy $\e{=}\e_p(b)$ 
and some choice of the amplitudes $A_j$ in eq.~(\ref{defP}) 
the normalized wave function 
of the relative motion vanishes at $r{=}b$ in all channels,
then the $P$-matrix has a pole at $\e_p(b)$. As shown in Ref.~\cite{P-orig}, 
its residue can be factorized:
\be
P_{ij}(\e)=\overline P_{ij}(\e)+\xi_i{~}\frac{r}{\e-\e_p}{~}\xi^T_j{~}.
\lb{pole}
\ee
We choose the vector $\xi$ to be normalized: $\sum_{i=1}^{n}\xi_i^2=1$.
Obviously, the converse statement is also valid: if at some energy $\e_p$
the $P$-matrix has a pole then one can find the amplitudes $A_j$ in 
eq.~(\ref{defP}) such that
\be
\frac{\psi_i(b)}{||\psi||}=0\qquad\forall i=1,\ldots,n~.
\lb{psi=0}
\ee
 
Now we are ready to explore the connection between the poles of the 
$P$-matrix and 
the quark-bag calculations. Remember that for now $b$ is taken to be 
larger than the range of the strong forces. In this case we just saw that 
the $P$-matrix poles 
(primitives) $\e_p(b)$ occur at the energies at which the
relative wave function of the two hadron system
vanishes at $r=b$.  We claim that
these are just the eigenenergies, $\e_n(R)$, of the multi-quark system that
has the quantum numbers of the two-hadron 
\begin{figure}
\fpsxsize=1.5in
\def\fpsangle{90}
\centerline{\fpsfile{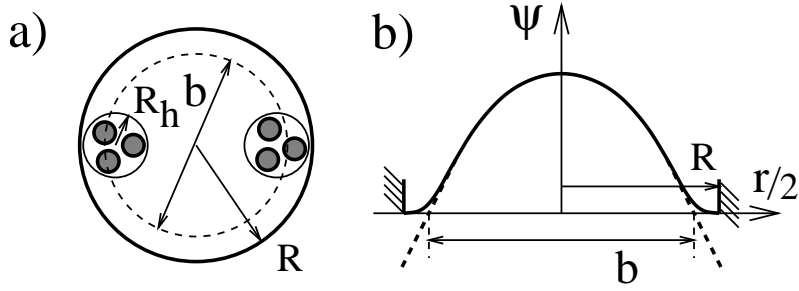}}
\medskip\medskip
\caption{
(a) A two-baryon system is confined in a hypothetical spherical shell with a 
radius $R$. 
(b) Consider a state of this system with 
a definite energy $\e_n$ (the ground energy in the figure).
The wave function of the centers of the $3q$-subsystems 
(solid line) strictly vanishes at the shell boundary. At the same 
energy $\e_n$, the wave function of the baryons
that are not constrained by the shell (dashed line) 
vanishes when their relative separation equals
$b=2R-2R_h$.}
\label{fig1}
\end{figure}    
\medskip
\noindent
system and is
confined in a hypothetical spherical shell\footnote{We require that the 
center of mass of the system always stays at the shell center. 
To realize it ``technically'' one can imagine a massless shell.} 
with a radius $R(b)$.
The radius $R$ is approximately half of $b$.
In fact, if we imagine that the hadrons are constrained in their
motion so that the matter density, $\rho({\bf r})$, vanishes
when $|{\bf r}-{\bf R}_{cm}|\ge b$~,
then the wave function of 
their relative motion $\psi(r)$ vanishes at  
\be
b=2R-2R_h{~},
\lb{Rbh}
\ee     
were $R_h$ plays the role of the hadron radius,
as shown in Fig.~\ref{fig1}.

We see that for a large value of $b$ there is one-to-one correspondence
between the $P$-matrix poles and the eigenenergies of a physical
system which is put into a hard-wall shell. Now let us make $b$
smaller. The $P$-matrix, as defined by eq.~(\ref{PS}), will preserve
a pole structure (see eq.~(\ref{pole})) 
but the parameters $\e_p$, $r$, and $\xi$
will change with $b$.  If $b$ goes to $b'$ the related
$P$-pole shifts to $\e'_p$, satisfying the equation
\be
\e'_p=\e_p-\xi^T\frac{r}{\overline P(\e'_p)+k'_p{~}\cot{~}k'_p\Delta b}
{~}\xi 
\qquad\mbox{where}\qquad \Delta b = b'-b {~}.
\lb{eb}
\ee
The new residue $r'$ and the channel couplings $\xi'$ can be easily
expressed in terms of~$\e'_p$.\footnote{The application of the 
identity eq.\rf{inv} and some algebra yield:
$\xi' r' \xi'^T{=}R{~}\xi r \xi^TR^T$
with $R(\e'_p,\Delta b)\equiv \frac{k'_p}{\sin{(k'_p{\Delta}b)}}{~}
\frac{1}{\overline P(\e'_p)+k'_p\cot{(k'_p{\Delta}b)}}$~.} 
Taking a small variation of $b$ one gets the differential equations:
\be
\frac{\partial\e_p}{\partial b} =-r
\lb{r}
\ee
and
\be
\frac{\partial}{\partial b}{~}(\xi r \xi^T)=
-{~}\xi r\xi^T \overline P -\overline P\xi r\xi^T 
\equiv  -{~}\{\xi r\xi^T,\overline P\}{~}.
\lb{R-der}
\ee 
The last equation can be also presented as
\be
\left.\overline P_{ij}(\e,b)\right|_{\e=\e_p}=-\frac{1}{2r} 
                 \frac{\partial r}{\partial b} \xi_i\xi_j
                 -\xi_i \frac{\partial \xi_j}{\partial b}
		 -\frac{\partial \xi_i}{\partial b} \xi_j+
		{~}\overline{ \overline P}_{ij}(\e_p,b)
\lb{Po}
\ee
where the matrix $\overline{ \overline P}$ is orthogonal to the vector $\xi$~:
$\overline{ \overline P}_{ij}{~}\xi_j=\xi_j{~}
\overline{ \overline P}_{ji}=0$.

It was noted by M.~Soldate$^{\cite{Sold}}$ 
that decreasing the radius of the shell in Fig.~\ref{fig1}
imposes additional constraints on the system inside and, 
therefore, causes the eigenenergies of its states, $\e_n(R)$, to grow. 
By eq.\rf{r}, $r>0$ is a necessary condition to have 
\be
\frac{\e'_p-\e_p}{b'-b}<0~.
\ee
It is also the sufficient condition if the matrix 
${\partial \overline P(\e)}/{\partial \e}$ 
is negative semidefinite, in particular if $\overline P(\e)$
is a constant, that can be shown from eq.~(\ref{eb}).

When the shell radius reaches the size of a few Fermis we should 
treat the system inside as a single quark-bag rather than two hadrons,
but strong interaction invalidates eq.~(\ref{defP}) 
at such distances. Then it becomes difficult$^{\cite{Sim2}}$  
to relate the system eigenenergies, $\e_n$, to the position of the $P$ poles,
$\e_p$. Nevertheless, we might expect that there is a size of the shell 
$R_0$ when the quark-gluon system inside is already simple enough for
our theoretical tools, while $\e_n(R_0)$ and $\e_p(b_0)$ are still close 
to each other. 
It was proposed in Ref.~\cite{P-orig} that at this $R_0$ 
the quark system in the shell may be treated as a single bag\footnote{
For a two-nucleon system, {\it e.g.} deuteron, $R_0$ must be taken
at much larger radius of the pion exchange forces, and this system
is better described by the pion exchange potential rather 
than the quark-bag model.} 
and its eigenstates
can be calculated in  perturbative QCD with current quark masses. 
This assumption reflects the idea that the bag interior is a phase built up on 
the perturbative vacuum. Alternatively, it could be a phase in which chiral 
symmetry is spontaneously broken, yielding constituent quarks with 
renormalized couplings and pion-like  excitations. Finally, one might also 
attempt to exploit lattice methods.
Anyway, if we estimate the position of the $P$-matrix poles $\e_p(b_0)$ and 
their orientation in the channel 
space $\xi_i(b_0)$, eqs.~(\ref{r}) and (\ref{Po}) will provide us with 
other ingredients of the $P$-matrix.
\subsection{Calculation of P}

In principle, {\it all} the information about the $P$-matrix 
can be rigorously obtained from calculations involving only
hadronic sizes. To this end one should solve the quark dynamics
and parameterize the hadronic wave function according to eq.~(\ref{defP}).
The external interaction can be taken into account as described in 
Ref.\cite{P-orig}.  In the absence of powerful methods applicable to scales of
order 1 $fm$ we resort to bag model phenomenology. 
To specify $P(\e)$ we require the primitive 
energies, $\e_p$, the residues, $r^{(p)}$, 
the channel coupling vectors, $\xi_i^{(p)}$, and the nonsingular 
part, $\overline P_{ij}(\e)$. 
We treat each of them in sequence below.

The primitive energies were already considered in the previous subsection.
Let us remember that they were identified with the eigenenergies, $\e_n$, of a 
quark-gluon system subject to confining boundary conditions at a sphere 
$R(b)$. 
 
In the determination of the vectors $\xi^{(p)}$ 
it is important to take account of 
flavor symmetry. For example, one may consider $SU(3)_f$ when describing
$\La\La$ scattering or $SU(2)_f$ for the $np$ system. In the scattering of
two scalar mesons $SU(3)_f$ is badly violated and the $SU(2)$ isospin 
symmetry is more appropriate.  
If the flavor symmetry were exact, the mass of all hadrons 
belonging to one multiplet would be the same. The states of an interacting 
system confined by the shell would also form flavor multiplets.
The eigenenergies of the states in one multiplet would be equal, and
the $P$-matrix would be $SU(n_{f})$ symmetric, whatever
the size of the shell. We do not observe this in reality
because of the difference in the current quark masses. Nevertheless,
the smaller the shell, the better the coupling vector reflects the flavor
symmetry. Let us show this in specific examples.

{\it Baryon-Baryon:}
Imagine two $\Lambda$-particles inside a {\it macroscopic} spherical shell.
To be specific, suppose they are in the ground energy
state with $J=0$ and assume that the fusion of the $\Lambda$'s
into one $H$-dibaryon$^{\cite{H-orig}}$ is energetically forbidden, {\it
i.e.\/}
$M_H>2M_{\Lambda}$. For the macroscopic shell the $\Lambda{-}\Lambda$ 
interaction is negligible, and the ground state is unique with the eigenenergy
$\e_p \simeq 2M_{\Lambda}$. This $\Lambda\Lambda$ system belongs to
the symmetrized product of the two $SU(3)_f$ baryon octets that decomposes
into the following irreducible representations:
\be
(8\otimes 8)_{\rm sym}=27\oplus 8\oplus 1{~}.
\lb{8*8}
\ee
However the $\Lambda\Lambda$ state can not be attributed to any of those 
irreducible parts, therefore the coupling vector $\xi_i$ in the $P$ pole 
corresponding to the ground state is {\it not} SU(3) symmetric.

Now we gently contract the shell so that the system remains
in its ground state. When the shell radius reaches
the order of 1~$fm$, the scale of the confinement starts to 
overcome the $s$-quark mass, and $SU(3)_f$ symmetry gradually
emerges. The $\Lambda$'s inside split into a ``gas'' 
of 6 strongly interacting quarks. 
Due to the color-magnetic interaction, the ground state of this
system now does occur at the flavor 
singlet$^{\cite{H-orig},\cite{B.W.}}$:\footnote{Modulo small $SU(3)_f$ 
violation due to current quark masses.}
\be
|H\rangle=\sqrt{\frac{1}{5}}{~}|BB\rangle+
\sqrt{\frac{4}{5}}{~}|\b 8 \cdot \b 8\rangle
\lb{Hdec}
\ee
where $|\b 8\cdot\b 8\rangle$ denotes a singlet superposition of
two color octet baryons and 
\begin{eqnarray}
|BB\rangle & = &\sqrt{\frac{1}{8}}{~}\left\{
|\Xi^-p\rangle-{~}|\Xi^0n\rangle+{~}|p\Xi^-\rangle-{~}|n\Xi^0\rangle
\right.\nonumber\\
& + &|\Sigma^-\Sigma^+\rangle+{~}|\Sigma^+\Sigma^-\rangle-
{~}\left.|\Sigma^0\Sigma^0\rangle+{~}|\Lambda\Lambda\rangle \right \}
\lb{BBsingl}
\end{eqnarray}
is the flavor singlet state composed of two physical baryons. 

In order to explore the interpretation of $\xi_i$ as the bag state orientation 
in the channel space, consider the parameter $b$ in eq.~(\ref{defP}) 
independently for each channel. 
If at the shell boundary the interaction is negligible,
the ``partial residue'' of the $i$-th channel will be
\be
r\xi^2_i=-\frac{\partial}{\partial b_i}\e_p
  \propto \left.\left|\frac{\partial\psi_i}{\partial r_i}
                \right|^2\right|_{r_i=b}{~}
\lb{pr}
\ee
($\psi$ is the normalized wave function of the confined system, obeying
$\left.\psi\right|_{r=b}{=}0$).
Thus $r\xi^2_i$ is associated with the ``partial pressure'' on the shell 
walls. Let us consider eq.\rf{pr} in the basis where the singlet 
state given by eq.\rf{BBsingl} is one of the basis vectors. 
Then for a small $b$ we have 
$\xi_i\simeq\delta_{i,singlet}$~, that is the residue 
$\xi_i r \xi^T_j$ of the lowest $P$-matrix pole is almost a $SU(3)_f$ singlet. 
As a first approximation we can take the vector $\xi$ corresponding to the
exact $SU(3)_f$ symmetry ({\it cf.} eq.~(\ref{BBsingl})): 
\be
\xi_i=\pm\sqrt{\frac{1}{8}}{~},{~~~~}
i=\Xi^-p,{~}\Xi^0n,{~}p\Xi^-,{~}n\Xi^0,{~}\Sigma^-\Sigma^+,{~}\Sigma^+\Sigma^-,
{~}\Sigma^0\Sigma^0,{~}\Lambda\Lambda{~}.
\lb{xi}
\ee

{\it Meson-Meson:}
The $f_0(980)$ resonance has the quantum
numbers $I(J^{PC})=0{~}(0^{++})$ and decays strongly into
$\pi\pi$ and $\bar{K}K$. Because of the great difference between the $\pi$ and
$K$ masses, it is not
realistic to assume $SU(3)_f$ symmetry even within the confinement
radius.  The $SU(2)_f$ symmetric decomposition for
$f_0$ reads :
\be
|f_0\rangle=\alpha_K\sqrt{\frac{1}{4}}{~}\left\{|K^-K^+\rangle-{~}|\bar{K^0}K^0\rangle+
      |K^+K^-\rangle-{~}|K^0\bar{K^0}\rangle\right\}+
\lb{fdec}\\
 +\alpha_\pi\sqrt{\frac{1}{3}}{~}\left\{|\pi^-\pi^+\rangle+{~}|\pi^+\pi^-\rangle-
      {~}|\pi^0\pi^0\rangle\right\}+\alpha_\eta{~}|\eta\eta\rangle+{~}\alpha_{c}{~}|c\rangle 
  \nonumber
\ee
where $|c\rangle$ stands for confined channels, {\it e.g.} 
glueball, and $\Sigma|\alpha_i|^2=1$. Therefore, the $P$-matrix for $\pi\pi$
scattering has a pole around $980{~}MeV$, and its ``orientation'' in
the channel space $\xi$ is given by the normalized projection of the 
decomposition~(\ref{fdec}) onto the two-particle channels $\pi\pi$, 
$\bar{K}K$, and $\eta\eta$.

Without a deeper understanding of confinement, we are only able to provide 
crude estimate for the dynamical parameters $r$ and $\overline P$.
These will serve as a guide in the next sections.  Ref.~{\cite{P-orig}} 
contains rather visual reasoning concerning the residue $r$
that we paraphrase in the present context as follows. 
Let us consider the ``partial pressure'' 
$p_i$ on the shell walls due to the $i$-th flavor component of the system: 
\be
4\pi R^2p_i\equiv -{~}\frac{\partial \e_p}{\partial R_i}{~}.
\lb{pp}
\ee 
We suppose that there is a radius of the shell $R_0$ when $p_i$ can 
both be calculated perturbatively in the quark-bag model and
attributed to the hadrons in the $P$-matrix approach. 
In the two-baryon example above the pressure exerted by the $\Lambda{-}
\Lambda$ subsystem is
\be
4\pi R^2 p_{\Lambda\Lambda} = 
-\frac{\partial \e_n}{\partial R}{~}\lambda{~}\xi^2_{\Lambda\Lambda}
{~~~~~}\mbox{ with }{~~}\lambda=\frac{1}{5}
{~~}\mbox{ and }~~\xi^2_{\Lambda\Lambda}=\frac{1}{8}{~},
\lb{p1b} 
\ee 
as given by the $SU(3)_f$ symmetric bag model, 
eqs.~(\ref{Hdec},\ref{BBsingl}). 
In the $P$-matrix formalism it is 
\be
4\pi R^2 p_{\Lambda\Lambda} = -{~}\frac{\partial b}{\partial R}{~}
  \frac{\partial \e_p}{\partial b_{\Lambda\Lambda}}{~}
  \xi_{\Lambda\Lambda}^2=
  \frac{\partial b}{\partial R}{~}r{~}\xi_{\Lambda\Lambda}^2{~},
\lb{p1p} 
\ee 
see eqs.~(\ref{pr},\ref{xi}). Comparing the right hand sides of
eqs.\rf{p1b} and\rf{p1p} we get
\be
r\simeq\frac{\partial R}{\partial b}{~}
  \frac{\partial \e_n}{\partial R}\left.\lambda{~}\right|_{R=R_0}
{~}.
\lb{r-est}
\ee
The important result is that the residue
$r$ is suppressed by the factor $\lambda{<}1$ with 
respect to its natural scale.

If one estimates primitive masses and residues in the bag model, 
the radius of the confining shell, $R_0$, is predetermined 
by the bag model virial theorem as     
\be
R_0\simeq 5M^{1/3}{~}GeV^{-1}{~},
\lb{R0}
\ee 
where $M$ is the mass of the quark-bag state measured in $GeV$.
In the discussion above we assumed that close to the boundaries of
the hypothetical shell in Fig.~\ref{fig1} the quark-gluon matter 
behaves as two almost non-interacting hadrons.
Let these hadrons have comparable masses, $M/2$, and 
treat them as separate rigid quark bags with the
radius given by the formula\rf{R0} as
\be
R_h\simeq 5\left(\frac{M}{2}\right)^{1/3}{~}GeV^{-1}\simeq 
   \frac{1}{2^{1/3}}~R_0 \simeq 0.8 {~}R_0{~},
\lb{Rh}
\ee
so that from eq.\rf{Rbh}
\be
b_0 \simeq 2R_0 - 2R_h \simeq 0.4{~}R_0{~}.
\lb{bR1}
\ee  
The assumption that the hadrons of the size $2R_h\simeq 1.6 R_0$ do not 
interact in the shell of the diameter $2R_0$ is absurd, and
the result\rf{bR1} should be considered as a lower bound on $b_0$.
An upper bound can be found from the opposite extreme
when the quarks inside the cavity are assumed to be 
uncorrelated with each other. This was done in 
the original paper$^{\cite{P-orig}}$ by Jaffe and Low. 
They chose $b_0$ that approximately matched the density of the 
free hadron-hadron 
wave function, $\psi(r)\propto \sin(\pi r/b_0)$,
vanishing at $r=b_0$, to the density of the centers of mass 
of two-quark clusters (for mesons), when all the quarks in the cavity 
moved independently.
For a meson-meson system that yielded
\be
b_0\simeq 1.4{~}R_0{~},
\lb{bR}
\ee 
and should give a smaller result for three-quark baryons.
From this discussion we conclude that the radius in eq.\rf{R0} is
too small for reliable calculation of the $P$-matrix but may suffice
for estimates of the order of magnitude, with $b_0\sim R_0$.

We can say even less about the matrix $\overline P(\e)$.
Definitely, it has poles corresponding to the other bag states.  
Eq.~(\ref{Po}) suggests that in the interstitial region
\be
\overline P_{ij}\sim \frac{1}{b_0}~.
\lb{Pbar_est}
\ee
\section{Corresponding S-matrix}

Now we turn our attention to the quantity connected to actual scattering 
experiments -- the $S$-matrix.
In the Subsection 3a we express $S$ and its singularities in terms of the
$P$-matrix discussed earlier. Then we consider in detail the threshold 
effects and their interference with primitives.

\subsection{General equations}
 
In the previous section we argued that the poles of the $P$-matrix
have fundamental significance. Taking $P$ in the pole form, eq\rf{pole},
\be
P_{ij}(\e)=\overline P_{ij}(\e)+\xi_i{~}\frac{r}{\e-\e_p}{~}\xi^T_j{~},
\lb{pole1}
\ee
one can easily reconstruct the corresponding $S$-matrix from
eq.~(\ref{SP}). In the denominator of eq.~(\ref{SP}) one has to deal with 
the inversion of a matrix having the structure 
$A_{ij}+\xi_{i} a \xi^T_{j}$ and 
the following identity comes handy
\be
\frac{1}{A+\xi a \xi^T}=\frac{1}{A}-\frac{1}{A}{~}\xi
  \frac{a}{1+\xi^T{\displaystyle\frac{a}{A}}{~}\xi}
  \xi^T{~}\frac{1}{A}{~}.
\lb{inv}
\ee
After some calculations we obtain
\be
S_{ij}=\overline S_{ij}-i\chi_{i}{~}\frac{1}{\e-\e_r+
  i{~}{\displaystyle\frac{\gamma}{2}}}
       {~}\chi_{j}{~}.
\lb{Sgen}
\ee
In this formula $\overline S$ is the background or potential scattering 
produced by 
$\overline P(\e)$:
\be
\overline S(\e)=e^{-ikb}{~}\frac{{\frac{1}{\sqrt{k}}} \overline P
  {\frac{1}{\sqrt{k}}} +i} {{\frac{1}{\sqrt{k}}} 
   \overline P {\frac{1}{\sqrt{k}}}-i}{~}e^{-ikb}{~}.
\lb{SoP}
\ee
Eq.\rf{SoP} also enables one to find $\overline P(\e)$ when the potential
scattering, $\overline S(\e)$, is known.
The pole term in eq.~(\ref{Sgen}) has diverse manifestations
in cross-sections, that are discussed in the next subsection. As shorthand
for them we will make free use of the word ``resonance'' recognizing that
a true resonance occurs only under somewhat limited conditions.
The ``resonance'' couplings to the hadronic channels $\chi_i$ are 
\be
\chi_i(\e)=\sqrt{2r}{~}e^{-ik_ib}\sqrt{k_i}
 \left(\frac{1}{\overline P-ik}\right)_{ij}\xi_j{~}.
\lb{chiP}
\ee
For the energy dependent ``resonance'' position and the width in the 
denominator of eq.~(\ref{Sgen}) we have
\be
\e_r(\e)-i{~}{\displaystyle\frac{\gamma(\e)}{2}}=
  \e_p-\xi^T\frac{r}{\overline P-ik}{~}\xi{~}.
\lb{esP}
\ee
The real and imaginary parts in eq.~(\ref{esP}) can be easily separated.
In order to do it, we write the many-channel momentum matrix $k$ as
\be
k=q+i\kappa{~},
\ee
where $q$ and $\kappa$ are real and refer to the open and closed channels
correspondingly. Recalling that for the strong interaction $\overline P$ is 
also real, we find
\be
\e_r(\e)=\e_p-{~}\xi^T
  \frac{r}{\overline P+\kappa+q
  \frac{1}{\overline P+\kappa}q}{~}\xi{~}
\lb{erP},
\ee
\be
\gamma(\e)=2r{~}\xi^T
 \frac{1}{\sqrt{1+\left(\frac{1}{\overline P+\kappa}
  q\right)^2}}{~}\frac{1}{\overline P+\kappa}{~}q{~}
 \frac{1}{\overline P+\kappa}{~}\frac{1}{\sqrt{1+\left(q
  \frac{1}{\overline P+\kappa}\right)^2}}{~}\xi{~}.
\lb{gP}
\ee
These equations are valid for a nonsingular $\overline P{+}\kappa$ and 
an arbitrary $q$.
One can write the total width in eq.~(\ref{gP}) as a sum of partial width 
$\gamma_i$ over only the open channels: 
\be
\gamma=\sum_{\stackrel{\mbox {\scriptsize open}}{\mbox
{\scriptsize channels}}}\gamma_i{~}
\lb{sumg}
\ee
with the $i$-th partial width:
\be
\gamma_i=2rq_i\left(\frac{1}{\overline P+\kappa}{~}
  \frac{1}{\sqrt{1+\left(q
  \frac{1}{\overline P+\kappa}\right)^2}}{~}\xi
 \right)_i^2{~~~~~}.
\lb{gi}
\ee

As expected, the elements $S_{ij}$ in eq.~(\ref{Sgen}) between the
{\it open} channels form a unitary sub-matrix:  
\be
S^{(0)}S^{(0)\dagger}=1{~},
\lb{unitarity}
\ee
where $S^{(0)}$ stands for the physical scattering matrix that is the 
restriction of $S$ to the open channels. 
Note that the unitarity of the physical $S$-matrix does not impose any 
additional restrictions on the $P$-matrix poles and residues or
on the  $\overline S(\e)$~, except that the physical part of 
$\overline S$ be unitary by itself. As long as $P$ is hermitian,
$S$-matrix unitarity is automatically taken care by eq.~(\ref{SP}). 

As we saw earlier, $P(\e,b_0)$ is completely determined by the dynamics in the 
microscopic domain where the interaction is strong, and it is not influenced
by the region in the configuration space where the system is represented
by two freely moving hadrons. Thus, {\it all kinematical effects are 
absorbed in eqs.~(\ref{Sgen}-\ref{gP})}. We proceed to study them next.
\subsection{Threshold analysis}

At a threshold the kinematics plays a key role. Threshold singularities
of the $S$-matrix and their analytical structure are well known and 
conveniently described in the $K$-matrix parameterization. 
The $P$-matrix formalism should provide 
similar results but allows a dynamical point of view, however.
In fact, we argue that at a threshold the reduced $K$-matrix 
and it's poles also have a  quantitative {\it dynamical} interpretation.
The object of this section is to give a simple classification scheme for
threshold effects and present an account of observed cross-sections 
for those cases. Let us emphasize that although threshold phenomena by 
themselves are well investigated, a quark state nearby may produce some 
deviations from the ``standard'' picture.

The analytical structure of $S(\e)$ at a threshold is complicated by
its many-sheeted structure with a branch point at the threshold energy.
The usual correspondence between $S$-matrix poles and physical states becomes 
more subtle. In fact, it is known$^{\cite{Pening-book}}$ that two different 
kind of poles may appear. In Ref.~\cite{Pening-book} Pennington points out 
that a bound or a virtual state near a threshold produced by a long-range 
potential gives rise to only one pole in the $S$-matrix. 
On the other hand, a tightly bound 
multi-quark state results in a pole on each energy sheet, and at a threshold, 
where two sheets merge, the two poles are equally important. 
We incorporate this observation into a general scheme. 
The $S$-matrix singularity analysis will provide a systematic arrangement of 
physical phenomena that to a good extent is independent of specific 
experimental circumstances.

Suppose the $P$-matrix has a pole at the energy $\e_p$ in the vicinity of 
a threshold. 
It might be the $H$-dibaryon at $\Lambda\Lambda$ 
threshold ($2230{~}MeV$) or $f_0(980) / a_0(980)$ at $\bar{K}K$ 
threshold ($990{~}MeV$).
The $S$-matrix in eq.~(\ref{Sgen}) 
has a pole at the energy $\e_s$ when the denominator $\e{-}\e_r(\e){+}i
\gamma(\e)/2$ vanishes.  Taking $\e_r{-}i\gamma/2$ from eq.\rf{esP} we get the 
the following equation for the $S$ poles generated by the quark state:
\be
\e_s=\e_r(\e_s)-i{~}{\displaystyle\frac{\gamma(\e_s)}{2}}=
     \e_p-\xi^T\frac{r}{\overline P(\e_s)-ik(\e_s)}{~}\xi {~}.
\lb{eseq}
\ee
The right hand side of eq.\rf{eseq} is a multivalued 
function because the momentum
\be
k(\e)= {\mbox{diag}}(k_1, \ldots, k_n) =
       {\mbox{diag}}\left(\sqrt{{~}2m{~}(\e-\e_{\rm th1})},
       \ldots, \sqrt{{~}2m{~}(\e-\e_{\rm th{~}n})}\right)
\lb{ke}
\ee
has branch points at all threshold energies $\e_{\rm th{~}i}$ .
We will call the channel with its threshold close to the $P$ pole the 
``singular'' one and the other channels, correspondingly, ``non-singular''.
A pole in $S(\e)$ disturbs the cross-section if only its position on
the complex energy Riemann surface is close enough to the physical region, 
which is the side of the real energy cut where each $k_i$ is either real and 
positive (for open channels) or 
$k_i=i\kappa_i$ with real and positive $\kappa_i$ (for closed channels). 
Accordingly, we are interested in the solutions of eq.~(\ref{eseq})
for which all non-singular channel momenta are taken to be close to the 
positive real or upper imaginary semi-axis. As for the singular momentum 
branches, each  of them is important provided the corresponding solution 
of eq.\rf{eseq} is close enough to the threshold. The natural measure of 
being ``close enough'' on the $k_{singular}$ complex plane is $1/b_0$,
as will be seen in the course of our work. That is when we reckon only one
channel as the singular one at our energy range, we have assumed that 
in all the other coupled channels the momenta, perhaps imaginary,  are 
larger than $1/b_0$ in absolute value.

Let us suppose that at the threshold under consideration all 
the non-singular channels are closed and label  the singular one with 
the subscript $i{=}1$. This condition is applicable to the $\La{}\La$ 
threshold in the previous baryon-baryon example or for $N{}N$ 
scattering. It is not true for the $f_0(980)$ or $a_0(980)$ resonances
near $K\bar K$-threshold because the non-singular $\pi\pi$ channel is open.
In the general case, discussed at the end of the section, 
our equations below remain unchanged but most of the parameters become complex.
In the following work we are neglecting the energy dependence in
the background term ${\overline P}$ and in the non-singular momenta 
$k_{i\not=1}$. As for ${\overline P}$, it must be a good 
approximation inside the complex circle 
\be
\left|k_1 b_0\right| < 1{~},
\lb{valcirc}
\ee
if we suppose that no other $P$ poles occur in this energy range.
The energy interval corresponding to eq.\rf{valcirc} is
around $20~MeV$ for $\La{}\La$ and $60~MeV$ 
for $K\bar K$ scattering. In these and 
many other cases $k_{i\not=1}\simeq const$ is also valid in  
the most of the range (\ref{valcirc}). 

Eq.\rf{eseq} for the $S$ poles energy can be rewritten in terms of 
the related momentum in the singular channel $k_{1s}$ 
($\e_s=k_{1s}^2/2m+\e_{th1}$) as
\be
\e_{th1}+\frac{k_{1s}^2}{2m} = \e_p+\Delta\e_p-i~\frac{k_{1s}\rho_1 c_1 / 2m}
 {c_1-ik_{1s}}{~},
\lb{kleq}
\ee
where $\Delta\e_p$~, $\rho_1$~, and $c_1$ are the following parameters
\be
\Delta\e_p \equiv -\xi^T\frac{r}{\C}{~}\xi{~},{~~}
\rho_1 \equiv 2mr \left[\left(\frac{1}{\C}\right)_{1i}\xi_i\right]^2{},{~~}
c_1 \equiv \left[\left(\frac{1}{\C}\right)_{11}\right]^{-1}{~},
\lb{param} 
\ee
and
\be
\bar\kappa_l \equiv \left\{{\displaystyle
    \begin{array}{cl}
       \frac{1}{i}k_l{~},&l\not=1{~};\\
       0{~},& l=1{~}.
\end{array}}
\right.
\lb{kpbar}
\ee
Notice that $\bar\kappa$ and the parameters in eq.\rf{param} are real since
we work with the lowest threshold ($\e_{th1}{<}\e_{th~l,~l\not=1}$), 
and they are approximately constants in the circle described by 
eq.~(\ref{valcirc}) where ${\overline P},\bar\kappa\simeq const$. 
In this approximation eq.~(\ref{kleq}) is cubic with respect to 
the unknown $k_{1s}$. That is we can find from zero to three $S$-matrix
poles inside the near-threshold circle, eq.~(\ref{valcirc}).

There is a natural physical interpretation for all the three possible
solutions of eq.~(\ref{kleq}) as well as for the parameters defined by
eq.~(\ref{param}).
First we consider the narrow resonance limit $r\rightarrow 0$ 
so that $\rho_1\rightarrow 0$ and $\Delta\e_p\rightarrow 0$. 
Then the solutions of eq.~(\ref{kleq}) are
\be
k_{1s}^{(1,2)}\simeq \pm \sqrt{2m(\e_p-\e_{th1})}{~},
  {~~}k_{1s}^{(3)}\simeq -ic_1{~}.
\lb{ksol}
\ee
This limiting case suggests that two of the three possible 
$S(\e)$ poles, namely $k_{1s}^{(1,2)}$, are generated by the 
quark state, one pole for each $k_1$ branch. The third solution
$k_{1s}^{(3)}$ is produced by the background $\overline P$ rather 
than the quasi-bound quark state. In the approximation of eq.~(\ref{ksol})
it is clear because $c_1$ is determined only by the 
background part of the $P$-matrix. This $S$-matrix pole will still be
present and located exactly at $-ic_1$ when there is no quasi-bound quark 
state at all and the $P$-matrix is given by only its background part:
$P=\overline P$. We want to call to mind that if $c_1$ is estimated at the 
threshold then the solution $k_{1s}^{(3)}{\simeq}-ic_1{~}$ is reliable
only when $|c_1b_0|<1$.

For an arbitrary $r$ it is very fruitful 
to consider the reduced $R$-matrix of the elastic scattering in the first 
channel defined as
\be
R^{(red)}_1(\e,b) =  \frac{1}{{\displaystyle P^{(red)}_1(\e,b)}}
\equiv \left(ik_1 \frac{e^{2ik_1{}b}S_{11}(\e)+1}{e^{2ik_1b}S_{11}(\e)-1}
\right)^{-1}
=\frac{1}{k_1}\tan(k_1b+\delta_1^{(elastic)})
\lb{Rdef}
\ee
({\it cf.} eq.~(\ref{PS})). In other words, $R^{(red)}_1$ is the
inverse logarithmic derivative of the first channel radial wave function 
$\left.\psi_1/\psi'_1\right|_{r=b}$, provided the incident wave is also taken
in the first channel.
After some algebra $R^{(red)}_1$ can be expressed in terms of the same 
parameters (\ref{param}), making their phenomenological interpretation 
especially transparent:
\be
R^{(red)}_1 = \frac{1}{c_1}-\frac{\rho_1/2m}{\e-(\e_p+\Delta\e_p)} 
\lb{Rpole}
\ee
Thus $c_1$ is responsible for the non-resonance (background) elastic 
scattering in the singular channel, whereas $\rho_1$ determines the resonance 
strength reduced to the first channel\footnote
{\label{rho_eff}
From the form of eq.\rf{Rpole} one might conclude that ${-}\rho_1/2m$ is
the residue of the pole in $R^{(red)}_1(\e)$. This is not correct in general
because $\Delta\e_p$ may depend on energy and the actual residue is 
$\left.{-}\rho_1^{eff}/2m=-\frac{\rho_1/2m}{(1-\frac{d\Delta\e_p}{d\e})}
 \right|_{\e=\e_p+\Delta\e_p}$. This effect is especially important
at a threshold were the energy dependence in $\Delta\e_p$ is strong.
A similar remark applies to $c_1$~.}, 
and $\Delta\e_p$ is naturally 
associated with the ``hadronic'' shift in its energy. 
Particularly, the resonance turns 
into a bound state when, and only when, $\e_p+\Delta\e_p<\e_{th1}$.

Now the classification of a near-threshold system behavior is 
straightforward. For example, let us look at the phase of the elastic 
scattering in the first, singular, channel. By the definition of 
$R^{(red)}$ (see eq.~(\ref{Rdef})), the phase equals
\be
\delta_1^{(elastic)}=\arctan{\left(k_1 R^{(red)}_1\right)}-k_1b_0{~}.
\lb{phase}
\ee 
and it experiences rapid variations in the threshold region $\left|k_1 
b_0\right| < 1$ when either of the two terms in eq.~(\ref{Rpole}) 
is large with respect to $b_0$.

First, suppose the second, resonant, term is negligible  or absent so that
$R^{(red)}_1 \simeq 1/c_1$. Then one gets the familiar scattering length
parameterization of the elastic amplitude
\be
f_{11} \equiv \frac{1}{k_1\cot\delta_1-ik_1}=
            \frac{1}{-\frac{1}{\overline a_1}-ik_1+O(b_0k_1^2)}{~}
\lb{Tbg}
\ee
with the scattering length
\be
\overline a_1 = b_0 - \left.\frac{1}{c_1}~\right|_{\e=\e_{th1}}{~}.
\lb{a}
\ee
If $\left|1/c_1\right|>>b_0$, the scattering length is anomalously large,
as it happens, for example, in $NN$ scattering. 
In this case the amplitude has a pole
at $k_1\simeq i\frac{1}{\overline a_1} \simeq -ic_1$ and it is 
either a bound state, such as the deutron, or a virtual state 
depending on the sign 
\begin{figure}
\fpsxsize=2in
\def\fpsangle{90}
\centerline{\fpsfile{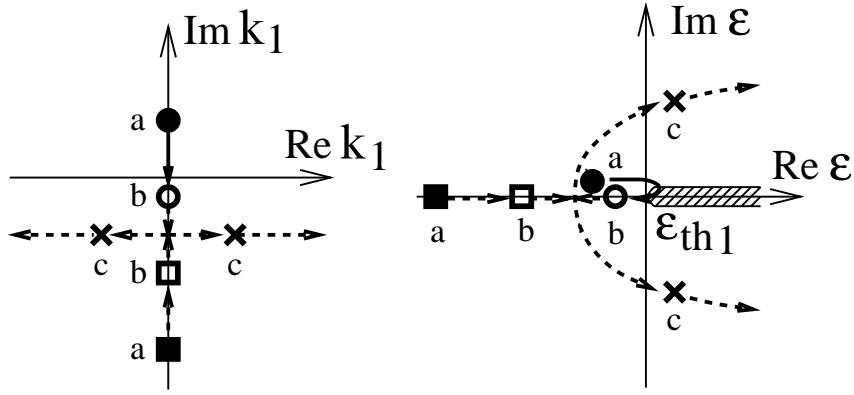}}
\medskip\medskip
\caption{
The $S$-matrix pole dynamics for a narrow quark state 
with its mass increasing at the lowest threshold.
The complex planes of the momentum $k_1$ (left) and 
the energy (right) are shown. Note that as the pole marked 
by the circle goes from the upper $k_1$ half-plane down to the 
lower half-plane, it moves from the physical energy 
sheet under the cut onto the unphysical sheet.}
\label{fig2}
\end{figure}
\medskip
\noindent
of $\overline a_1$~. Note that $O(b_0k_1^2)$
(effective range) corrections to the amplitude in eq.\rf{Tbg} come from both 
$c_1$ energy dependence and the $k_1b_0$ term in eq.\rf{phase}.

Second, we consider the situation when the resonant term in $R^{(red)}_1$
has its pole near the threshold and suppose that the ``background'' $1/c_1$ is
of the ``normal'' magnitude, $\left|1/c_1\right|\le b_0$, so that 
$(k_1/c_1)^2$ is much less than one
in the range of our interest, eq.\rf{valcirc}. If the width,
$\rho_1$, is large ($\rho_1 \ge 1/b_0$), we get the same ``scattering
length'' phenomenology with the scattering length
\be
a_1=\overline a_1-\left.\frac{\rho_1}{2m\e_1}~\right|_{\e=\e_{th1}},
~~~~~~\e_1 \equiv \e_p + \Delta\e_p - \e_{th1}
\lb{a1} 
\ee
and $O(b_0k_1^2)$ effective range corrections, just as in eq.\rf{Tbg}.
For such a large $\rho_1$ the scattering amplitudes and $S$-matrix can have 
again no more than one near-threshold pole that arises when 
$\left|\frac{\rho_1}{2m\e_1}
\right|>>b_0$ and is located on the imaginary $k_1$ axis:
\be 
k_{1s}\simeq i\frac{1}{a_1}\simeq -i\rho_1/2m\e_1~.
\lb{large-rho-pole}
\ee 
It will be a stable particle if 
$\e_p{+}\Delta\e_p{<}\e_{th1}$ and a virtual state if 
$\e_p{+}\Delta\e_p{>}\e_{th1}$. 
In the alternative case of 
a narrow resonance ($\rho_1{<}{<}1/b_0$) another $S$-matrix
pole appears. In fact, in the region $(k_1/c_1)^2{<}{<}1$ the 
equation \rf{kleq} that specifies the $S$ pole momenta becomes quadratic:
\be
\frac{k_{1s}^2}{2m^*}+i\frac{k_{1s}\rho_1}{2m}-\e_1=0{~},
     {~~~}m^*\equiv \frac{m}{1-\frac{\rho_1}{c_1}}{~}.
\lb{kqeq}
\ee
If $\rho_1 {<}{<} 1/b_0$ then $m^*\simeq m$, 
for we assumed that $\left|1/c_1\right|\le b_0$, and the solutions of 
eq.\rf{kqeq} have their average at
\be
\frac{1}{2}(k_{1s}^{(1)}+k_{1s}^{(2)})\simeq-i\frac{\rho_1}{2}{~}.
\ee
Therefore when $\rho_1 {<}{<} 1/b_0$ and the quark state is not far from the 
threshold, $\abs{\e_1}<b_0^{-2}/2m$, both solutions get into the 
circle\rf{valcirc}. The arrangement of the $S$ poles on the momentum and 
energy complex planes for different values of $\e_1$ is shown 
in Fig.~\ref{fig2}. Despite the intricate analytic structure, 
elastic scattering in the singular channel shows nothing more than than a 
narrow resonance slightly distorted by the threshold (see Fig.~\ref{fig7}~(a)
for example). 
The scattering phase will have the form 
$\delta_1^{(elastic)}=\overline\delta_1+\delta_1^r$ with
\be
\overline\delta_1 = \arctan{({k_1}/{c_1})}-k_1 b_0
                    \simeq -{\overline a_1}k_1
\ee
and
\be
\delta_1^r \simeq -\arctan{\left(\frac{k_1\rho_1/2m}
                         {\e-\e_p-\Delta\e_p}\right)}~,
\lb{phase_el}
\ee
omitting the terms $O(k_1^2/c_1^2)$ and $O(\rho_1/c_1)$.
In practice, the parameterization of scattering amplitudes by the reduced 
$R$-matrix, eq.\rf{Rdef}, is inconvenient since the parameter $b_0$ is
not strictly defined and model dependent. The reduced $K$-matrix,
$K^{(red)}_1(\e)\equiv \left.R^{(red)}_1(\e,b)\right|_{b=0}$ is free
from this deficiency and can be easily found:
\be
K^{(red)}_1(\e) \simeq - \overline a'_1 - \frac{\rho'_1/2m}
                 {\e-\e_{th1}-\e'_1}~,
\lb{Kred}
\ee 
where
\be
\overline a'_1 = \frac{\overline a_1}{1-\rho_1b_0}~,~~
\rho'_1 = \rho_1~\frac{1-\rho_1b_0+2m\e_r\overline a_1b_0} 
   {(1-\rho_1b_0)^2}~,~~
\e'_1 = \frac{\e_1}{1-\rho_1b_0}~.
\ee
The reduced $K$-matrix in eq.\rf{Kred} gives the correct elastic amplitude
\be
f_{11} = \frac{1}{\left(K^{(red)}_1\right)^{-1}-ik_1}
\ee
up to terms $O(k_1^2b_0c_1^{-1})$ regardless of the magnitude of $\rho_1$.

We have not considered the possibility that both terms in the 
reduced $R$-matrix in eq.\rf{Rpole} are
anomalously large. This situation may give rise to many interesting
phenomena. Nevertheless, we will not discuss them here
because the chance of such a double accident (large ``potential''
scattering length $\overline a_1$ and a quark state close to the threshold) 
should be small and we are not aware of a real two-hadron system with these 
properties.

So far we treated elastic amplitudes only. Inelastic $S$-matrix
elements also present practical interest, even when the other, non-singular,
channels are closed and inelastic scattering is energy forbidden. 
For example, if a six-quark singlet state $H$ exists
between $\La\La$ and $N\Xi$ thresholds, the following production experiment is
possible: $K^-d\rightarrow K^0\Xi N\rightarrow K^0H\rightarrow K^0\La\La$ 
where the intermediate $\Xi$ is virtual. With reasonable assumptions, 
the amplitude of this process is proportional to $S_{\La\La,\Xi N}$.
In general, the off-diagonal $S$-matrix element between the singular, first,
channel and a non-singular channel $l$ is given by the formula\rf{Sgen} as
\be
S_{1l}=\overline S_{1l}-i{~}\frac{\chi_{1}\chi_{l}}{\e-\e_r+
  i{~}{\displaystyle\frac{\gamma}{2}}}{~~},
\lb{Sinel}
\ee
where the ``resonance'' energy and width are completely determined by the 
same three parameters --  $\e_p{+}\Delta\e_p$, $\rho_1$, and $c_1$~:
\be
\e_r-i{~}\frac{\gamma}{2}=
 \e_p+\Delta\e_p - \frac{\rho_1c_1}{2m}{~}\frac{(k_1/c_1)^2}{1+(k_1/c_1)^2}
 -i{~}\frac{k_1\rho_1/2m}{1+(k_1/c_1)^2}{~},
\lb{es1}
\ee
and so is $\chi_{1}$~:
\be
\chi_{1}=\sqrt{k_1\rho_1/m}{~}\frac{e^{-ik_1b_0}}{1-ik_1/c_1}{~}.
\lb{chi1}
\ee
Near the first threshold when $k_1{<}{<}1/b_0,1/\overline a_1$ 
we have the anticipated result that $\gamma\propto k_1$~,
$\chi_{1}\propto\sqrt{\gamma}\propto\sqrt{k_1}$~, and $\e_r$~,
$\chi_{l}$ have finite magnitude~:
\be
\e_r-i{~}\frac{\gamma}{2}=\left( \e_p+\Delta\e_p+O(k_1^2) \right) -
         i~\frac{k_1\rho_1/m}{2}~\left(1+O(k_1^2)\right)~,
\lb{es1ap}
\ee
\be
\chi_{1} \simeq \sqrt{k_1\rho_1/m}{~}(1-i\overline a_1k_1)~~,
~~~~
\chi_{l}= \sqrt{k_l\rho_l/m}{~}(1+O(ik_1)){~}.
\lb{chil1}
\ee

 In conclusion, we discuss the new features introduced
to our analysis when one of the non-singular strongly coupled channels is
{\it open}. In this case all the previous equations of this section 
formally remain valid, of course with the subscript ``1'' replaced by 
the number of the singular channel, ``2''. The first real 
difference comes from the fact that $\kappa_l$ in the analogue of
eq.\rf{kpbar},
\be
\overline\kappa_l=\frac{1}{i}~k_l~(1-\delta_{l2})~,
\ee
is now complex for $l{=}1$  giving non-vanishing imaginary part 
to the parameters $\Delta\e_p$, $\rho_2$, and $c_2$ in eq.\rf{param}. As a 
consequence, the scattering lengths $a_2$ or $\overline a_2$ become complex.
At the second threshold similar to eq.\rf{es1ap}, 
\be
\e_r-i{~}\frac{\gamma}{2}=\left( \e_p+\Delta\e_p+O(k_2^2) \right) -
         i~\frac{k_2\rho_2/m}{2}~\left(1+O(k_2^2)\right)~.
\lb{es2ap}
\ee
But $\e_r$ and $\gamma$ were originally defined as real quantities whereas
$\Delta\e_p$ and $\rho_2$ in this equation are complex. Therefore the ``smooth
term'', $\e_p+\Delta\e_p+ O(k_2^2)$~, in eq.\rf{es2ap}\\
\begin{figure}
\fpsxsize=3in
\def\fpsangle{90}
\centerline{\fpsfile{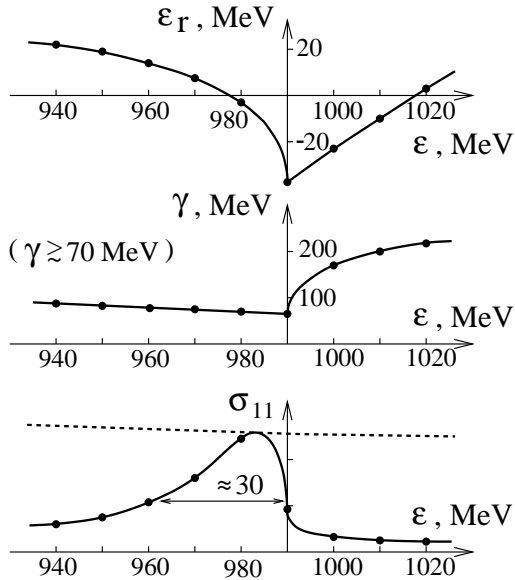}}
\medskip\medskip
\caption{
The effective resonance position ($\e_r$), effective width 
($\gamma$), and elastic cross section ($\sigma_{11}$) as functions of 
energy $\e$ for a two-channel model with a $P$-pole close to the
second threshold at $990{~}MeV$. The dashed line on the $\sigma_{11}$
plot is the unitary limit for $s$-wave scattering.
The half-width of the peak in the cross-section is 
considerably less than~$\gamma$.}
\label{fig3}
\end{figure}
\medskip
\noindent
contributes to both $\e_r$ and $\gamma$
providing a non-vanishing width just at the second 
threshold and below. Due to the complex $\rho_2$, the ``cusp term'', 
$ik_2\rho_2/2m~$, also contributes to $\e_r$ and to $\gamma$ above and under 
the threshold. Nevertheless, from the definition 
of $\rho_2$ one can see that its imaginary part is suppressed by the factor 
$b_0/k_1$, which is less than one by our earlier assumptions.
That is $\gamma(\e)$ still has the bigger cusp just above the 
threshold and $\e_r(\e)$ just below. In Fig.~\ref{fig3} 
we present a two-channel example where two hypothetical particles with the
equal masses $140~MeV$ in one channel and $495~MeV$ in the other are coupled 
by a primitive at $1040~MeV$, which is $50~MeV$ above the second threshold. 
The couplings $\xi_1$ and $\xi_2$ are taken to be 
equal. Notice that in this model $\gamma$ exceeds $70~MeV$ for
all energies in the range of interest
whereas the observed half-width of the corresponding resonance
is as small as $30~MeV$. 
This is the result of strong energy dependence 
in $\e_r(\e)$ and $\gamma(\e)$ at the threshold
that has another consequence. Namely, the right slope of the
resonance curve, which is closer to the threshold, is considerably
more steep than the left one. 
The masses in this example were chosen to be
those of pion and kaon that form the 
decay channels of the $f_0(980)$ resonance,
eq.\rf{fdec}. Of course, this simple nonrelativistic model is not able to
provide an adequate description of $f_0(980)$, but the resemblance of the 
cross-section in Fig.~\ref{fig3} to the observed $\pi\pi$ 
scattering is striking.
  
When the singular threshold is not the lowest in energy,
we also have the possibility of studying the elastic scattering 
in the {non-singular} open channel.  For a narrow quark state bringing two 
near-threshold poles to the $S$-matrix one can use 
the same formula\rf{Sgen} with the resonance position and width just 
discussed (see eq.\rf{es2ap} and the text that follows it). 
One pole, produced by a large ``background'' scattering length 
$\bar a_2$ or a broad quark state at the threshold, may also appear as 
a narrow resonance in the non-singular channel.
To see this, let us write down the element $S_{11}$ in full, extracting the 
rapidly varying singular momentum $\kappa_2 \equiv \frac{1}{i}k_2$:
\be
S_{11}=e^{-2ik_1b_0}~\frac{1+ik_1/d_1}{1-ik_1/d_1}
       ~\frac{1+\kappa_2/c_2^*}{1+\kappa_2/c_2}{~}.
\lb{S11}
\ee
We will present the expression for $d_1$ in a moment, and $c_2$ is defined as 
earlier in eqs.(\ref{param},\ref{kpbar}) with $1{\rightarrow}2$ 
but $\overline P$ is now identified with $P$.
In particular, $1/c_2$ gives the scattering length in the singular 
channel: 
\be
a_2=b_0-1/c_2~.
\lb{a2}
\ee
Separating the real and imaginary parts in $1/c_2$ one obtains
\be 
\frac{1}{c_2}=\frac{1}{d_2}\left[1-\alpha_{12}\frac{k_1^2}{k_1^2+d_1^2}\right]
   +i~\frac{\alpha_{12}}{d_2}~\frac{k_1d_1}{k_1^2+d_1^2}{~}.
\ee 
The quantities $d_1$, $d_2$, and $\alpha_{12}$ are real and constructed
from the matrix 
\be
D_{ij}\equiv P_{ij} + \sum_{l\not=1,2} \delta_{il} \delta_{jl} \kappa_l
  ~~~~~~~~(i,j=1,2,\dots, n)
\ee
as follows:
\be
\frac{1}{d_k}
 \equiv \left(\frac{1}{D}\right)_{kk}{~~~~}(k=1,2)~,~~~~~~
    \alpha_{12} \equiv d_1d_2~\left(\frac{1}{D}\right)_{12}
    \left(\frac{1}{D}\right)_{21}{~}. 
\ee
If for some system $Re(1/c_2)$ is negative and $Im(1/c_2)$ turns out
to be small then the S-matrix element in eq.\rf{S11} has a sharp resonance
when
\be
\kappa_2~Re{\left(\frac{1}{c_2}\right)}\simeq -1~.
\lb{r_cond}
\ee 
This effect has a natural interpretation. In fact, the small imaginary part
of $1/c_2$~, which by eq.\rf{a2} up to the sign equals the imaginary part of 
the scattering length in the second channel, means that the coupling
between the first and the second channel is weak. If we turned this 
coupling off completely, we would find that the hadrons in the second channel
form a bound state just at the energy satisfying to eq.\rf{r_cond}, similar to
the bound state at $k_{1s}^{(3)}$ in eq.\rf{ksol}. The small coupling
of this state to the first channel gives rise to the resonance in the first
channel elastic scattering exactly at the would-be-bound state energy 
determined by the equation\rf{r_cond}. Presumably, 
this is the origin of $\Lambda (1405)$ resonance in the $\Sigma\pi$ 
scattering just under the $\bar K N$ threshold.

\begin{figure}
\fpsxsize=2in
\def\fpsangle{90}
\centerline{\fpsfile{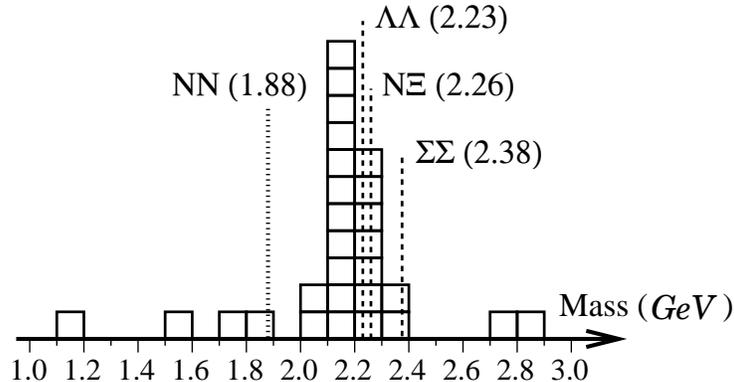}}
\medskip\medskip
\caption{The histogram showing the theoretical predictions of the 
$H$-dibaryon mass verses the energy scale and its weak -- $NN$ -- 
and strong -- $\La\La,~N\Xi,~\Sigma\Sigma$ -- decay thresholds.}
\label{fig4}
\end{figure}
\medskip

\section{Application to the H dibaryon}

\begin{figure}
\fpsxsize=1in
\def\fpsangle{90}
\centerline{\fpsfile{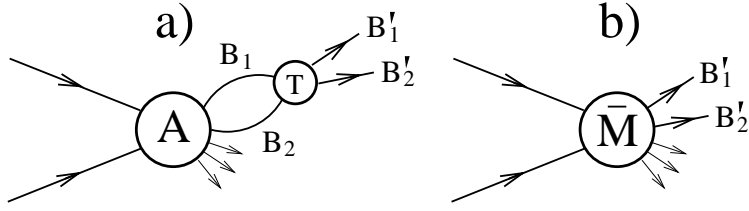}}
\medskip\medskip
\caption{
(a) An inclusive two baryon production amplitude sensitive to 
the $H$-particle coupling. $B_1B_2$ and $B'_1B'_2$ are some of the baryon 
pairs from eq.~(76) 
%<<<Careful !!!>>>
in the text. (b) The background amplitude for
$B'_1B'_2$ production.}
\label{fig5}
\end{figure}

The hypothetical $H$-dibaryon with its mass close to the $\La\La$ threshold
was our major illustration throughout this paper. About thirty theoretical 
predictions of its mass have been made. As shown in Fig.~\ref{fig4}\ct{H_hist},
they range from below $2M_N$, when the $H$ would be a more stable form 
of hadronic matter than nuclei, to $2.8~GeV$, well above its 
strong decay threshold into $\La\La$ at $2.23~GeV$. 
So far, the experimental searches for $H$ dibaryon have given 
inconclusive results\footnote{
The two candidate events reported in 
Ref.~\cite{Hcandidates} have recently been reanalyzed\ct{Brkh.memos} 
and found likely to be misidentified $K_L$ decays.}, 
although no definitive observations of double hypernuclei which would 
preclude a deeply bound $H$ have not been reported either.
These experiments were usually 
aimed at the $H$-particle which is stable to strong decay. 
However, the uncertain theoretical and experimental status 
of the $H$-dibaryon leaves much room for an $H$ which is heavier than 
$2M_{\La}$. In this case it would appear as a resonance-like
structure in the two-baryon sector with $S{=}{-}2$,
$I{=}0$, and $J{=}0$. Furthermore, the small projection of $H$ onto the
baryon channels, $1/5$ in the flavor $SU(3)$ limit by eq.\rf{Hdec},
suggests that the corresponding resonance might be more or less narrow
and could be detected in an experiment which is sensitive to
a two-baryon scattering amplitude in the $SU(3)_{f}$ singlet channel. 
The amplitude for such a typical production experiment
schematically shown on Fig.~\ref{fig5}~a) will be of the form
\be
M=\sum_{B_1B_2}A_{B_1B_2}T_{B_1B_2\to B_1'B'_2}(s')+\overline M~,
\lb{M}
\ee
\medskip
\noindent
where $A_{B_1B_2}$ and $\overline M$ are some smoothly varying 
production amplitudes and
the two-baryon scattering amplitude $T_{B_1B_2\to B'_1B'_2}(s')$
exhibits resonance behavior when the $B'_1B'_2$ invariant mass
$\sqrt{s'}$ is close to the mass of the $H$-particle. 
Let us remember that the flavor and the spin singlet
$H{=}(uuddss)_{singl}$ gives rise to a pole (primitive) in the two-baryon 
scattering $P$-matrix that by eq.\rf{xi} has equal couplings, 
$\xi_i=\pm\sqrt{1/8}$, to the following channels
\be
\Xi^-p,{~}\Xi^0n,{~}p\Xi^-,{~}n\Xi^0,{~}\Sigma^-\Sigma^+,{~}\Sigma^+
\Sigma^-,{~}\Sigma^0\Sigma^0,{~}\La\La~.
\lb{channels}
\ee
The two-baryon threshold energies are shown in Fig.~\ref{fig4}.
In the sections~IIb,~IIIa we saw that the interaction with hadrons effectively
downshifts the H particle energy by the amount $\Delta\e_p$ estimated
in the Appendix as 
\be
\Delta\e_p \sim (-40~MeV) - (-150~MeV){~}.
\ee

A deeply bound $H$,
\be
\e_p+\Delta\e_p<\e_{\La\La}-b_0^{-2}/2m\simeq 2210 ~MeV
\ee 
obviously would not affect the baryon scattering. If it is unbound
and far from hadronic thresholds, 
\be
\abs{\e_p+\Delta\e_p-\e_{\La\La/N\Xi/\Sigma\Sigma}}>b_0^{-2}/2m\simeq 20~MeV~,
\ee
and the residue $r$ given by eq.\rf{r-est} is sufficiently small,
the $H$ should appear as a typical resonance\rf{Sgen} in 
baryon-baryon scattering that has comparable couplings
$\chi_i$, eq.\rf{chiP}, to all of the channels\rf{channels} which are open 
at the resonance energy. 
However, whether the $H$-resonance should be narrow is an open question.
Let us characterize the $H$ width by 
$\Delta E_H {\equiv} b_0^{-1}\rho_{\La\La}/2m$ that is
the energy interval where the resonant term in 
eq.\rf{Rpole} is important for $\La\La$ scattering. Then our estimates in
the Appendix give that $\Delta E_H$ may range from $10~MeV$ to $40~MeV$.
Unknown nonperturbative dynamics may also change 
the situation significantly, and the resonance can be so broad that 
the two-baryon quasielastic amplitudes will not show any apparent peaks.
In this case the reduced $R$-matrix defined by eq.\rf{Rdef} in terms of  
the measurable elastic scattering 
phase will still have a pole as given by eq.\rf{Rpole}. 
Therefore, the $H$-primitive can be experimentally {\it observed} even 
if there is no actual particle associated with it.

\begin{figure}
\fpsxsize=4in
\def\fpsangle{90}
\centerline{\fpsfile{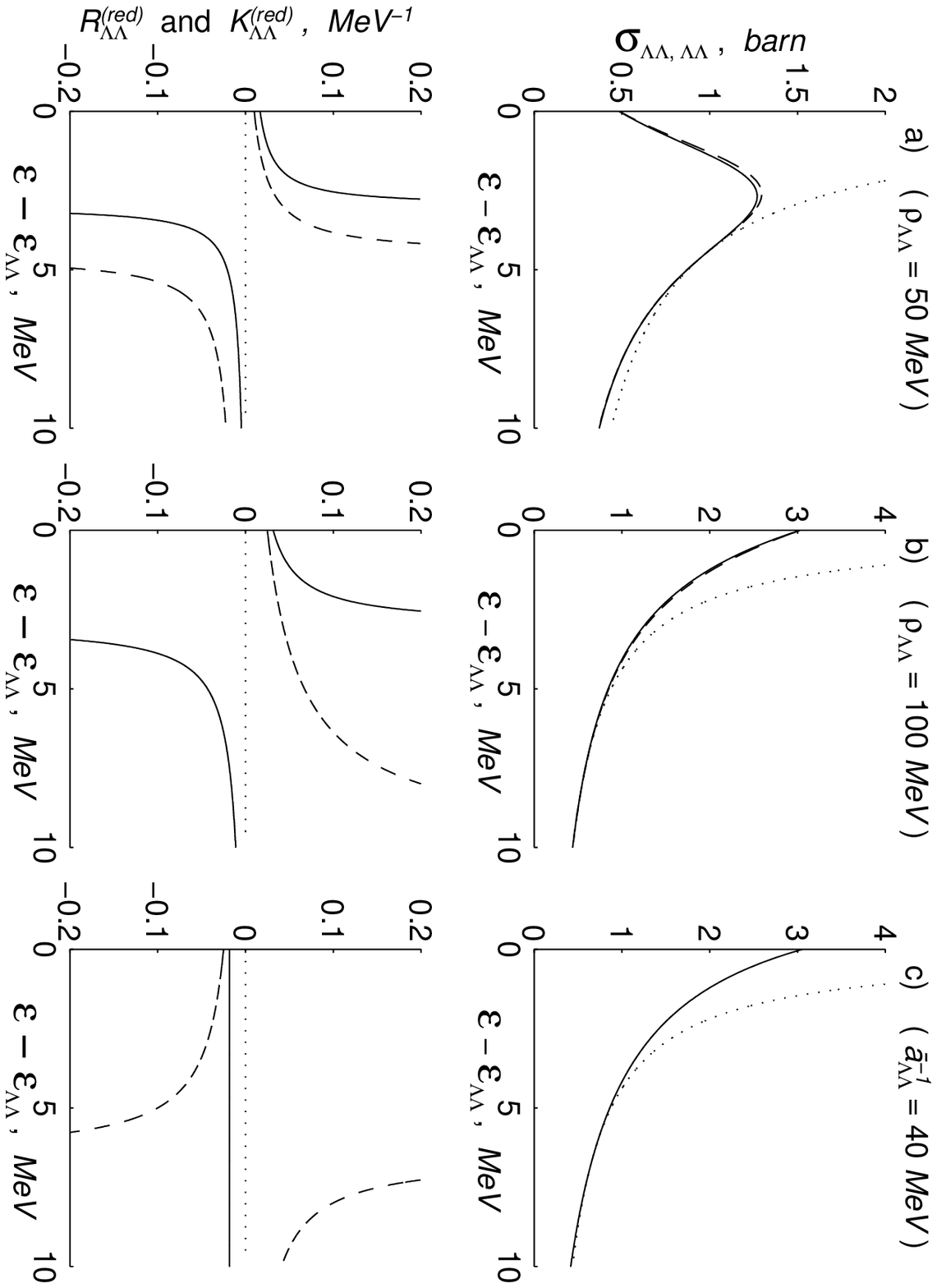}}
\medskip\medskip
\caption{The cross-section of the $s$-wave $\La\La$ elastic scattering 
(top) and the corresponding reduced $R$ (bottom, solid) 
and $K$ (bottom, dashed) matrices when the effective $H$-particle mass 
is close to the $\La\La$ threshold at $\e_{\La\La}=2230~MeV$: 
$\e_p+\Delta\e_p=\e_{\La\La}+3~MeV$. In the first two cases 
the parameter $a_{\La\La}^{-1}$ that characterizes the potential scattering
has a ``normal'' magnitude $200~MeV$ and the width parameter 
$\rho_{\La\La}$ is $50~MeV$ and $100~MeV$ for the plots (a) and (b) 
correspondingly. In the case (c) the reduced $R$-matrix does not have
a pole: $R^{(red)}_{\La\La} = c_{\La\La}^{-1} = const$ but 
it is anomalously large with the 
scattering length $\overline a_{\La\La}{=}(40~MeV)^{-1}$. 
The corresponding cross-section is almost indistinguishable
with the case (b). The dashed line on the top figures (a) and (b) gives 
the cross-section calculated from eqs.~(59) and~(60).
The dotted line shows the unitary
limit, $4\pi/k^2$.}
\label{fig7}
\end{figure}

Now we consider what happens if the $H$ is close to one of the 
thresholds in eq.\rf{channels}. Suppose it is the lowest threshold, $\La\La$. 
There is no particular reason to expect anomalously large potential
scattering in this channel\ct{A&D}. 
If the pole in $R_{\La\La}^{(red)}$ is narrow,
the elastic $\La\La$ scattering amplitude and phase will have the classical 
Breit-Wigner forms with the width proportional to the $\La\La$ relative 
momentum, eq.~(\ref{phase_el}), and the inelastic 
(quasielastic) scattering will be given by eqs.(\ref{Sinel}--\ref{chil1}). 
All these quantities can be expressed
in terms of the three real parameters
$(\e_p+\Delta\e_p)$, $\rho_{\La\La}$, and $\overline a_1$
for elastic scattering and the additional parameter 
$\rho_{N\Xi(\Sigma\Sigma)}$  for 
transition to or from another channel. Close to the $\La\La$
threshold $R_{\La\La}^{(red)}$ approaches the reduced $K$-matrix 
of the elastic $\La\La$ scattering and one should find a narrow pole in
the latter as well.  If the couplings 
$\rho_{\La\La/N\Xi/\Sigma\Sigma}$ and the residues of the 
$R_{\La\La}^{(red)}$ or $K_{\La\La}^{(red)}$ poles
are not small, the $\La\La$ scattering will be well described by the 
scattering length in eq.\rf{a1} with no dramatic resonance effects, but this 
length and the cross-section may be anomalously large when the corresponding
pole in the $S$-matrix, eq.\rf{large-rho-pole}, 
is close to the $\La\La$ threshold.
In many aspects this very last possibility will resemble a quasi-bound
two-$\La$ state. 
The characteristic plots of the amplitudes for the
cases above are presented in Fig.~\ref{fig7}. 

Our formalism does not make specific predictions for background 
(potential) scattering
parameterized by $\overline P$. One should expect that $\overline P(\e)$
has no poles in the region of interest. These poles might come from the 
excited $H$ states and from other, non-singlet, flavor 
representations of the six-quark system. 
One-quark excitation energy is of the order of $\pi/b_0\sim 500{~}MeV$. 
As the Table I of the Ref.~\cite{H-orig} shows, the other six-quark  
flavor multiplets have their ground states at the energies at
least $200{~}MeV$ higher than the flavor singlet mass because of 
color-magnetic interaction. Therefore, the matrix $\overline P$ in 
eq.~(\ref{pole}) should be a smooth function of energy in a broad 
interval ${\sim}100{~}MeV$  around the singlet pole.  
Let us briefly discuss some suggested or ongoing experiments
which might be sensitive to an unbound or resonant $H$.
 
{\it Nucleon-Nucleus collisions:} The production of strange particles
is plentiful when a nucleus is struck by a nucleon or another nucleus.
This has been exploited in the search for the stable $H$-particle in 
the experiments E888 and E896 at the AGS.
However, the same abundance of $\La$'s or $\Xi$'s will present 
a problem for a detection of the $H$-resonance because only a small fraction
of $\La\La$ or $N\Xi$ pairs will be produced close enough in space
and with their relative momenta small enough to interact via $H$-formation. 
We estimate this fraction from the $\La\La$ and $N\Xi$ coalescence rate
computed in Ref.\cite{Cole}. That gives the order of $10^{-3}$, making
it very hard to see a resonant $H$ above the background uncorrelated 
$\La\La$ or $N\Xi$ pairs.

{\it Elastic Secondary Scattering:} One can produce two $\La$'s or another
baryon pair in eq.\rf{channels} and hope to extract their elastic 
scattering amplitude from interaction in the final state, 
Fig.~\ref{fig5}~a) with $B_1B_2=B'_1B'_2$~. 
The total amplitude will have the 
form in eq.\rf{M} where $\overline M$ will be
in general large due to ``direct'' $\La\La$ production, 
Fig.~\ref{fig5}~b). 

\medskip

\begin{figure}
\fpsxsize=3in
\def\fpsangle{90}
\centerline{\fpsfile{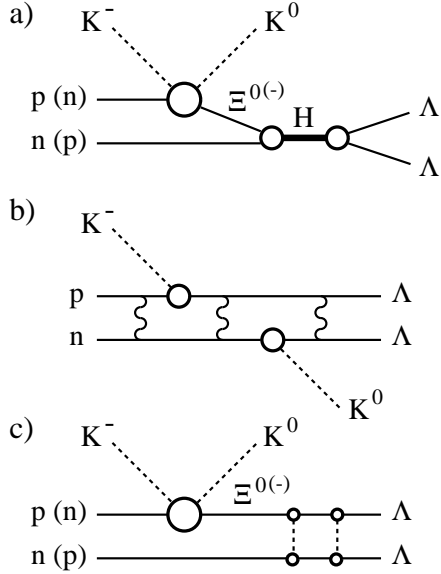}}
\medskip\medskip
\caption{
(a) The lowest-order processes in $K^-d\rightarrow K^0\La\La$ which
are sensitive to an $H$-resonance.
(b, c) The lowest-order background to this reaction.}
\label{fig8}
\end{figure}

{\it Quasi-Elastic Secondary Scattering:} We can also suppress the diagrams
like the one in Fig.~\ref{fig5}~b) by producing, 
{\it e.g.}, a virtual or real $N\Xi$ pair and then 
observing it scattering into $\La\La$. The reactions of this kind were 
considered by Aerts and Dover\ct{A&D} in connection with the 
stable $H$ production.
For example, take the process $K^-d\rightarrow K^0\La\La$. The diagrams that
we are interested in are shown in Fig.~\ref{fig8}~a). When $K^-$ and 
$K^0$ interact with different nucleons as in Fig.~\ref{fig8}~b), 
the final $\La$'s can be put on the mass shell only by some 
additional momentum exchange between them
that suppresses the corresponding amplitude and essentially makes it
higher order. In that order we also have soft meson exchange, e.g.
Fig.~\ref{fig8}~c). The cross-section for that was estimated in 
Ref.\cite{Macek} as
$\sigma_{(N\Xi\rightarrow \La\La)}v_{N\Xi}\simeq 10~mb$ at the $N\Xi$
threshold. If we take\\
\medskip
\noindent
$\left[\left(\frac{1}{\C}\right)\xi\right]_
{\La\La}\simeq \left[\left(\frac{1}{\C}\right)\xi\right]_{N\Xi}$, 
the formula\rf{Sinel} gives at the resonance peak, 
$\e{=}\e_r{\simeq}\e_{N\Xi}$,
that $\sigma_{(N\Xi\rightarrow \La\La)}v_{N\Xi}\simeq 60~mb$. 
Thus the background $\La\La$ production in this process may be 
comparable but does not exceed the resonance production 
as it did in the previous cases. The reactions of this form may
prove most promising for a search for a resonance $H$ above $\La\La$
threshold.

\section{Acknowledgments}
The authors would like to thank B.~Kerbikov and J.~Engelage
for stimulating discussions and useful references. 

\section*{Appendix}
\subsection*{The hadronic shift and width for the H-particle}

In the sections~IIb,~IIIa we saw that the interaction with hadrons effectively
shifts the H-particle energy by $\Delta\e_p$ given in 
eq.\rf{param}. We estimate this quantity and the other parameters,
$\rho_{\La\La}$ and $c_{\La\La}$, in eq.\rf{param} for the $H$-particle at the
$\La\La$ threshold. Altogether they compose the reduced 
$R$-matrix in the $\La\La$ channel (eqs.~(\ref{Rdef}-\ref{Rpole}))~:
\be
R^{(red)}_{\La\La} \equiv \frac{1}{k_{\La\La}}\tan(k_{\La\La}b_0+
  \delta_{\La\La}^{(elastic)})
  = \frac{1}{c_{\La\La}}-\frac{\rho_{\La\La}/2m}{\e-(\e_p+\Delta\e_p)}~. 
\lb{RLL}
\ee

Let us notice that if $\La$-particles did not interact with
each other, the $P$-matrix would have the lowest pole at 
$\e_p=2M_{\La}{+}\frac{(\pi/b_0)^2}{2m}$ with the residue
\be
r_0 = - \frac{\partial\e_p}{\partial b} = 
      - \frac{\partial}{\partial b}\left(\frac{(\pi/b)^2}{2m}\right) =
      \frac{\pi^2}{mb_0^3}~.
\lb{rfree}
\ee
For interacting $\La$'s one could expect that the residue $r$ 
is suppressed with respect to $r_0$ by the factor 
$\lambda{=}\frac{1}{5}$ due to the 
small projection of $H$ onto the baryon channels 
(see eq.~(\ref{r-est})).  On the other hand, if $r$ is computed in 
the bag-model then eq.\rf{r-est} and the virial theorem\ct{P-orig} give
\be
r_{bag~mod.} = \lambda \left.\frac{\partial \e_n}{\partial R}{~}
    \frac{\partial R}{\partial b}~\right|_{R=R_0} = \left.\lambda~
  \frac{3}{4}~\frac{M_H}{R}~\frac{\partial R}{\partial b}~\right|_{R=R_0}~
  \simeq \lambda~\frac{3}{4}~\frac{M_H}{b_0}~.
\lb{rbag}
\ee
Taking for $M_H$ the threshold energy $2M_{\La}$ and accepting that 
$b_0\simeq R_0 \simeq 5M_H^{1/3}{~}GeV^{-1}\simeq (150~MeV)^{-1}$, 
the ratio $\lambda r_0/r_{bag~m.}$ is 
$1:4$. This discrepancy is easily understandable. In fact, 
in eq.\rf{rfree} the pressure on the cavity walls that determines $r$
is exerted by heavy hadrons with the kinetic energy 
$(\pi/b_0)^2/2m<<M_H$ whereas in eq.\rf{rbag} we have
ultra-relativistic quarks and the pressure it is proportional to their
total energy $\e_{tot}\sim M_H$.

Assuming for simplicity that
$\overline P_{ij} \sim \delta_{ij} b_0^{-1} \sim \delta_{ij}{\cdot}150~(MeV)$,
we find that at the $\La\La$ threshold
\be
\rho_{\La\La}\sim 300~MeV~~,~~~~ \Delta\e_p\sim -150~MeV
\lb{pbag}
\ee
if we use the bag model prediction for $r$ in eq.\rf{rbag}, or
\be
\rho_{\La\La}\sim 70~MeV ~~,~~~~ \Delta\e_p\sim -40~MeV
\lb{pfree}
\ee
if $r$ is taken as $\lambda r_0$~. 
The contribution of the $\La\La$ channel, 
for which $\bar\kappa$ in eq.\rf{param} is zero, 
to $\Delta\e_p$ is comparable with the $N\Xi$ channels 
($p\Xi^-,~n\Xi^0,~\Xi^-p,~n\Xi^0$)
and the $\Sigma\Sigma$ channels give only $20\%$ to the total.
The actual values of the parameters $\rho_{\La\La}$ and $\Delta\e_p$ 
are probably somewhere in between the numbers in eqs.\rf{pbag} and \rf{pfree}.
Also notice that as described in the footnote on p.~\pageref{rho_eff},
the energy dependence in $\Delta\e_p$ may effectively change $\rho_{\La\La}$. 
For comparison, the hadronic shift computed by
Badalyan and Simonov\ct{BdSm} was found to be $15-30~MeV$ but they included
only coupling to the $\La\La$ channel, and larger values  
$150-200~MeV$ were obtained by B.~Kerbikov\ct{Kerb} and $100~MeV$ by
M.~Soldate\ct{Sold}.

\end{document}